\documentclass{ecai} 

\usepackage{latexsym}
\usepackage{amssymb}
\usepackage{amsmath}
\usepackage{amsthm}
\usepackage{booktabs}
\usepackage{enumitem}
\usepackage{graphicx}
\usepackage{color}

\usepackage{times}
\usepackage{soul}
\usepackage{url}
\usepackage{graphicx}
\usepackage{amsmath}
\usepackage{booktabs}
\usepackage{amssymb}
\usepackage{amsthm}
\usepackage{mathrsfs}
\usepackage{subfigure}
\usepackage{multirow}
\usepackage{algorithm}
\usepackage{algpseudocode}
\usepackage{xspace}
\usepackage{amsfonts}
\usepackage{float}
\floatname{algorithm}{Algorithm}
\newcommand{\sysname}{\texttt{SemiAdv}\xspace}
\newcommand{\et}{\textit{et al.\xspace}}

\usepackage{color}

\usepackage{soul}

\newtheorem{definition}{Definition}

\newcommand{\BibTeX}{B\kern-.05em{\sc i\kern-.025em b}\kern-.08em\TeX}

\begin{document}


\begin{frontmatter}


\paperid{1653} 


\title{\sysname: Query-Efficient Black-Box Adversarial Attack with Unlabeled Images}


\author[A]{\fnms{Mingyuan}~\snm{Fan}}

\author[B]{\fnms{Yang}~\snm{Liu}}

\author[A]{\fnms{Cen}~\snm{Chen}}

\author[C]{\fnms{Ximeng}~\snm{Liu}}


\address[A]{East China Normal University, China}

\address[B]{Xidian University, China}

\address[C]{Fuzhou University, China}


\begin{abstract}
    Adversarial attack has garnered considerable attention due to its profound implications for the secure deployment of robots in sensitive security scenarios.
    To potentially push for advances in the field, this paper studies the adversarial attack in the black-box setting and proposes an unlabeled data-driven adversarial attack method, called \sysname.
    Specifically, \sysname achieves the following breakthroughs compared with previous works.
    First, by introducing the semi-supervised learning technique into the adversarial attack, \sysname substantially decreases the number of queries required for generating adversarial samples.
    On average, \sysname only needs to query a few hundred times to launch an effective attack with more than 90\% success rate.
    Second, many existing black-box adversarial attacks require massive labeled data to mitigate the difference between the local substitute model and the remote target model for a good attack performance.
    While \sysname relaxes this limitation and is capable of utilizing unlabeled raw data to launch an effective attack.
    Finally, our experiments show that \sysname saves up to $12\times$ query accesses for generating adversarial samples while maintaining a competitive attack success rate compared with state-of-the-art attacks.
\end{abstract}

\end{frontmatter}


\section{Introduction}
\label{sec_introduction}
Convolutional neural networks have emerged as one of the most promising tools in various tasks such as object recognition, scene understanding, and even complex decision-making processes~\cite{robot1,robot2, fan2024guardian}.
However, plenty of recent studies point out the vulnerability of the networks to adversarial attacks~\cite{carlini2017towards, auto_pgd, fan2023enhance, surfree, chang2020restricted,fan2023trustworthiness,andriushchenko2020square}.
Adversarial attacks represent a sophisticated method of manipulating the output of a machine learning model by introducing carefully crafted, visually imperceptible noises to original inputs.
The vulnerability dramatically lowers the robustness of robots equipped with the networks in those application scenarios with high reliability and correctness requirements~\cite{chen2015deepdriving, das2018convolutional, yokota2020conformable}.
To further investigate the promising defense mechanism and alleviate the security concern, it is an essential task to build a groundwork of adversarial attacks for security evaluation.

There are two kinds of adversarial attacks: white-box attacks and black-box attacks.
White-box attacks allow the attacker to get the internal knowledge of the target model, including the model structure and training data~\cite{carlini2017towards, BIM, athalye2018synthesizing}.
However, while these attacks provide valuable insights into the vulnerabilities of models, the white-box assumption is too strong to be adopted in real-world applications.
In contrast, a more reasonable and ideal setting is only allowing the attacker to send (legitimate) inputs and receive corresponding outputs, which is called \textit{black-box adversarial attack}.
This restricted access poses significant challenges for the attacker but also reflects the conditions encountered in most real-world applications.

To promote practical attacks in the black-box setting, the previous work mainly follows two attack directions:
\begin{itemize}
    \item \textbf{Query-based attacks.}
    In this strategy, the attacker leverages gradient estimation to explore a perturbed input that can produce an approximate gradient to the original input~\cite{chen2017zoo, ilyas2018black, tu2019autozoom}.
    Techniques such as zeroth-order optimization or local search algorithms are employed to approximate the gradient using repeated queries to the model.
    Given a loose constraint for querying the target model, gradient estimation can achieve a high attack success rate.
    However, the high query budget limits its applicability for real-world applications. 
    \item \textbf{Transfer-based attacks.}
    Instead of directly attacking the target model, another popular attack method is finding a local substitute model that performs similarly to the target model, through which the attack setting is changed from ``black'' to ``white''~\cite{tramer2017space, activelearning, suya2020hybrid, fan2023robustness}.
    In detail, the attacker generates adversarial examples in the local substitute model and then harnesses them against the target model.
    The feasibility of such attacks is based on the transferability of adversarial samples.
    Moreover, due to the difficulty of narrowing the difference between the target and substitute models, transfer-based attacks usually require preparing massive labeled data for model estimation, thus yielding a low success rate.
\end{itemize}


This paper focuses on transfer-based attacks, and, to further relax the above constraints for conducting transfer-based attacks, we design \sysname.
\sysname distinguishes itself from previous work by effectively utilizing unlabeled raw data, which is easier to collect from open-source repositories, for launching attacks.
Furthermore, the only query access of \sysname is to ask the target model to label a few raw samples.
Other data used for training the substitute model are simply derived from these labeled samples by local semi-supervised learning.
Therefore, \sysname extremely lowers the hardness for local substitute model estimation and decreases the average query accesses required for black-box adversarial attacks.
Meanwhile, the query of \sysname behaves more ``normally'', i.e., making only one request for each attack input, rather than querying the identical input multiple times for local model fine-tuning as in previous works~\cite{carlini2017towards, tu2019autozoom, suya2020hybrid}.
As a result, \sysname presents a more practical way of launching black-box adversarial attacks.
Our contributions are three-folded:

%

\begin{itemize}
    \item 
    We explore a more realistic black-box setting with much stricter yet more realistic assumptions: 1) the query only returns the label, and 2) a tighter query budget with each sample to be only queried once. In such a setting, existing methods can hardly achieve satisfactory results based on our evaluations.

    \item We propose a novel yet effective adversarial attack, named \sysname, specifically designed for the black box setting.
    \sysname explores semi-supervised learning to utilize unlabeled data and overcomes the notoriously intensive query complexity for conducting the attacks.

    
    \item 
    We conduct extensive experiments on the benchmark datasets to verify the effectiveness of the proposed method.
    In all experiments, \sysname outperforms the state-of-the-art attack methods.
    Specifically, \sysname can achieve up to $12\times$ query complexity reduction with around 90\% attack success rate.
    
\end{itemize}

\section{Related Work}
\label{sec_related_work}

\begin{figure*}[!ht]
\centerline{\includegraphics[scale=0.5]{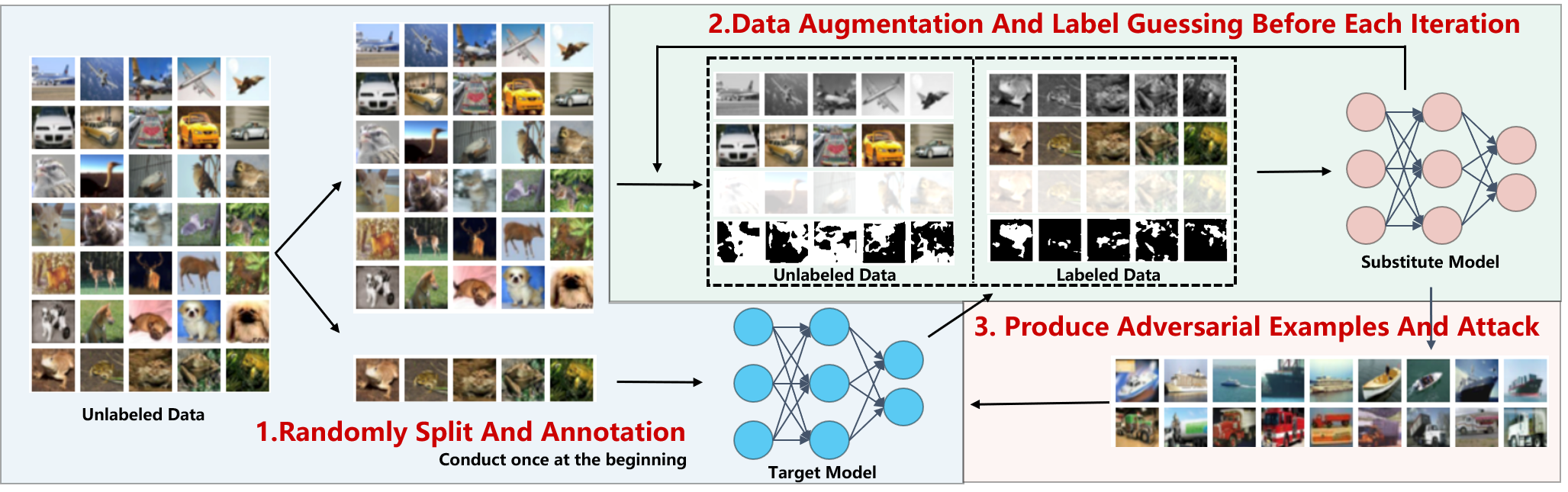}}
\caption{The workflow of \sysname.}
\label{fig_workflow}
\end{figure*}

\subsection{Adversarial Attack}

An essential observation about the adversarial attack was that the adversarial samples were transferable~\cite{tramer2017space}.
Instead of directly attacking the remote target model, the transferability allowed the adversary to launch white-box attacks on a local substitute model and use the locally generated adversarial examples to launch attacks~\cite{suya2020hybrid}.
Moreover, another popular adversarial attack was gradient estimation~\cite{chen2017zoo, tu2019autozoom}.
Chen \et~\cite{chen2017zoo} first introduced this type of attack in 2017, which was called ZOO. 
It combined finite-difference method and dimension-wise estimation methods to estimate gradients and utilized the famous Carlini-Wagner (CW) white-box attack~\cite{carlini2017towards} to find adversarial samples.
Although the follow-up work AutoZOO~\cite{tu2019autozoom} improved ZOO, its query number for gradient estimation is still too massive to apply in applications, e.g., over 15K queries per image for attacking ImageNet.
There are also other types of adversarial attacks that also attract lots of attention.
To attack the audio system, Alzantot~\et~\cite{alzantot2019genattack} proposed a genetic programming method with the prediction scores of the target model.
In $\mathcal{N}$Attack, Li \et~\cite{li2019nattack} redefined the adversarial sample search problem to be identifying a specific distribution that tends to be adversarial from a random space.
For these attacks, they broaden the potential directions for adversarial attacks but, limited by their applicable ranges, are not very practical in more general scenarios.

\subsection{Semi-Supervised Learning}

Semi-supervised learning leverages unlabeled data to enhance model performance.
\sysname harnesses a range of semi-supervised learning techniques to effectively minimize the number of queries required, including consistency regularization, entropy minimization, as well as proxy label.
Consistency regularization \cite{dualstudent,interpolation} promotes the consistency of the trained model's predictions across given training data and their perturbed versions.
Entropy minimization \cite{entropymini,Realistic} enforces the model to make confident predictions on unlabeled data regardless of the output label.
Proxy label \cite{selftraining,labelpropagation} instead uses the trained model to assign proxy labels to the unlabeled points, and these proxy labels are then used as targets.
In Section \ref{sec_approach}, we elucidate how \sysname exploits these techniques to achieve superior performance.

\section{Problem Formulation}
\label{sec_motivation}
In this paper, we hope to design a general black-box adversarial attack method that can be applied in variable application scenarios.
Based on the definition of adversarial attack (Definition \ref{def_adv}), we illustrate the improvements that \sysname aims to make.


\begin{definition}
\label{def_adv}
    (Adversarial Attack) Given a valid input image $x\in[0, 1]^{(w, h)}$ and a target neural network $f_{\theta}$, an effective adversarial attack $\mathcal{A}$ aims at exploring an optimized perturbation $\delta\in[0, 1]^{(w, h)}$ on $x$ such that $f_{\theta}$ gives a mislead output on $x + \delta$.
\end{definition}

Assume the legitimate label space is $T = \{y_0, y_1, ..., y_k\}$ and the correct label of $x$ is $y\in T$.
The optimization problem of adversarial attack can be formulated as follows.
\begin{equation}\label{eq_untargey_AE}
\begin{aligned}
    &\underset{\delta}{\text{minimize}}\quad \mathcal{L} = f_{\theta}(x + \delta, y^*) + \lambda D(\delta)\\
    & s.t.\quad (x + \delta)\in [0, 1]^{(w, h)}, \|\delta\|_{\infty} \leq \varepsilon,
\end{aligned}
\end{equation}
where $y^*$ is an arbitrary label $y^*\neq y$, $y^*\in T$, and $f_{\theta}(x + \delta, y_j)$ signifies the loss for perturbing $f_{\theta}$ to give a wrong output label, $D(\delta)$ is the distance between the true image $x$ and perturbed image $x + \delta$, $\lambda$ controls the influence of the perturbation and $\|\cdot\|_{\infty}$ denotes the $\ell_{\infty}$ norm item.


Further, for most black-box adversarial attacks~\cite{brunner2019guessing, chang2020restricted, andriushchenko2020square}, several constraints are introduced in real-world applications, which include: 1) unknown model structure and hyperparameters; 2) only access to model outputs (usually prediction scores); 3) limited access to the in-distribution training data; 4) a limited budget for target model query.
\sysname strengthens the last two constraints to make the attack stronger as follows.

\textbf{Unlabeled Data Only.} 
Besides assuming limited access to in-distribution training data, \sysname assumes that the prepared attack data are unlabeled, which can be collected by the attacker easier.

\textbf{Limited Query Access.} 
Except for a tight budget for querying the target model, \sysname limits each sample to be only queried once during the attack procedure.
Such constraint is introduced because, for most zero-order gradient based attacks, the attacker has to query the target model with the \textit{same inputs} successively until reaching a point where the adversarial images are generated successfully, e.g., the famous C\&W attack~\cite{carlini2017towards}.
Such an abnormal attack behavior can be simply detected by counting the query frequency for the same input in a short period.
Moreover, instead of the prediction score, \sysname only needs to access label outputs of the target model, which is more realistic, but harder to attack.


    


\section{Our Attack}
\label{sec_approach}
In this section, we present our proposed query-efficient black-box adversarial attack method, as demonstrated in Fig.~\ref{fig_workflow}. 
To the best of our knowledge, \sysname is the first to achieve the unlabeled data based black-box adversarial attack via semi-supervised learning, where both labeled and unlabeled data is fully utilized with query-efficient access.
Specifically, \sysname comprises two steps, i.e., substitute model training and local adversarial attack, which are outlined in Algorithm~\ref{protocol_model_training} and Algorithm~\ref{protocol_white_box_attack}, respectively.

\begin{algorithm}[!h]
  \caption{Substitute Model Training}
  \label{protocol_model_training}
  \begin{algorithmic}[1]
    \Require
    Unlabeled data set $D$;
    the target model $M$;
    the substitute model $M'$ and its parameter $\theta$;
    maximum epochs $J$;
    batch size $B$;
    label sharpening temperature $T$;
    learning rate $\alpha$;
    data augmentation ratio $N$;

    \State Initialize by selecting a random subset from $D$ and querying $M$ to form initial labeled set $D_0 = \{(x,y)|x \in D, y = M(x)\}$ and remaining set $D_1 = D \setminus D_0$.

    \For{$j \gets 1$ to $J$}
      \State Sample a batch of samples $\mathcal{X,U}$ from $D_0$ and $D_1$.
      \For {$b\gets 1$ to $B$}
        \State $\hat{x}_b=Augment(x_b)$.
        \For {$i\gets 1$ to $N$}
          \State $\hat{x}_{t,i}=Augment(x_t)$.
        \EndFor
        \State  Calculate the average prediction of the substitute model $\bar{p}_t=\frac{1}{N}\sum_{i=1}^{N}M^{'}(\hat{x}_{t,i})$.
        \State Sharpen the averaged prediction $p_t=Sharpen(\bar{p}_t,T)$.
      \EndFor
    
      \State Let $D^{'}_{0}=\{(\hat{x}_b,q_t)\}$ and $D^{'}_{1}=\{(\hat{x}_{t,i},p_t)\}$.
      \State $S=Shuffle(Concat(\mathcal{D}^{'}_{0},\mathcal{D}^{'}_{1}))$.
      \State $D^{'}_{0}=\{Mixup(D_{0,i}^{'},S_{i}) | 1 \leq i \leq D_{0}^{'}\}$.
      \State $D^{'}_{1}=\{Mixup(D_{1,i}^{'},S_{i+|D^{'}_{0}|}) | 1 \leq i \leq D_{1}^{'}\}$.

      \State $\mathcal{L}_{label} = \frac{1}{|\mathcal{D}_0'|} \sum_{(x, p)\in \mathcal{D}_0'} \mathcal{L}_{cross}(p, \mathcal{M}'(x))$.
      
      \State $\mathcal{L}_{unlabel} = \frac{1}{|\mathcal{D}_1'|} \sum_{(x, p)\in \mathcal{D}_1'} \|p - \mathcal{M}'(x)\|^2_2$.

      \State Update: $\theta \gets \theta - \alpha \nabla_{\theta}(\mathcal{L}_{label} +  \mathcal{L}_{unlabel})$.

    \EndFor
    
    \State \textbf{Return} the substitute model $M^{'}$.
    
  \end{algorithmic}
  
\end{algorithm}


\subsection{Substitute Model Training}

Due to the limited access to the labeled training data, it is essential to augment the data for substitute model training.
This step determines how similarly the substitute model performs to the target model, which further affects the attack success rate.
\sysname augments the data based on the semi-supervised learning technique. 
The detailed procedure of training a substitute model in \sysname is as follows.

1. \textit{Initialization.}
Assume the unlabeled data to be $\mathcal{D}$ and the target model to be $\mathcal{M}$.
The attacker randomly selects a subset of $\mathcal{D}$ and query $\mathcal{M}$ to form the initial attack data pool $\mathcal{D}_0 = \{(x, y)|x\in \mathcal{D}, y = \mathcal{M}(x)\}$ where $y$ is the label output by the target model.
We denote the remaining unlabeled data to be $\mathcal{D}_1$, which satisfies $\mathcal{D}_0 \cup \mathcal{D}_1 = \mathcal{D}$.
Note that this step is only performed once during the whole attack process, and \sysname only has to query each $x\in \mathcal{D}$ once.

2. \textit{Data Augmentation.}
After $\mathcal{D}_0$ is fed to the local substitute model $\mathcal{M}'$ (see Step 4), the attacker conducts data augmentation.
Inspired by semi-supervised learning methods~\cite{dualstudent,interpolation}, \sysname performs data augmentation, e.g., resize and crop, to expand both labeled and unlabeled data ($\mathcal{D}_0$ and $\mathcal{D}_1$).

3. \textit{Label Guessing.}
For each augmented unlabeled sample, \sysname guesses its label by first computing:
\begin{equation}
	\bar{p}_t = \frac{1}{N} \sum_{i = 1}^{N} \mathcal{M}'(\hat{x}_{t, i}),
\end{equation}
where $\bar{p}_t$ is the averaged prediction score of the $t_{th}$ unlabeled sample in $\mathcal{D}_1$,  $\hat{x}_{t, i}$ is the $i_{th}$ augmented sample of $x_t\in \mathcal{D}_1$ and $N$ is the total number of augmented samples based on the unlabeled sample.
Based on entropy minimization, a sharpening function is applied to the averaged prediction scores for reducing label distribution entropy.
\begin{equation}
	p_{t} = Sharpen(\bar{p}_{t}, T) =  \bar{p}_{t, k}^{\frac{1}{T}} /  \sum_{j = 1}^{K} \bar{p}_{t, j}^{\frac{1}{T}}.
\end{equation}
Assume the learning task to be $K$-classification and $k\leq K$.
$ p_{t, k}$ is the guessing prediction of the $t_{th}$ unlabeled sample.
$T$ is the temperature that controls the sharpening degree.

Moreover, \sysname adopts the mixup strategy proposed in~\cite{zhang2017mixup} to form the label and unlabeled sets for the next iteration of substitute model training.
Denote the union of augmented labeled set $D_0' = \{(\hat{x}_t, q_t)| q_t = \mathcal{M}(x_t)\}$ and unlabeled set $D_1' = \{(\hat{x}_{t, i}, p_t)| p_{t} = Sharpen(\bar{p}_{t}, T) \}$ as $\mathcal{D}' = \mathcal{D}_0' \cup \mathcal{D}_1'$.
After $\mathcal{D}'$ is shuffled, the $mixup$ function works on labeled data as follows.

We first select (sample, prediction) pairs from $\mathcal{D}'$, i.e., $(x, p)\in \mathcal{D}'_0$ and $(\hat{x}, \hat{p})\in \mathcal{D}'$. We then compute the following equation.
\begin{equation}
	\zeta = max(\eta, 1 - \eta),
\end{equation}
\begin{equation}
	x' = \zeta\cdot x + (1 - \zeta)\cdot \hat{x},
\end{equation}
\begin{equation}
	p' = \zeta\cdot p + (1 - \zeta)\cdot \hat{p},
\end{equation}
where $\eta$ is sampled from beta distribution that is determined by hyperparameter $\alpha$.
Finally, all elements in $\mathcal{D}_0'$ can be replaced with $(x', p')$ by repeatedly performing the above computations.
$\mathcal{D}_1'$ can also be updated similarly.

4. \textit{Local Model Training.}
With $\mathcal{D}_0'$ and $\mathcal{D}_1'$, the substitute model is updated based on the following loss function.
\begin{equation}
\begin{aligned}
    \mathcal{L} &= \mathcal{L}_{label} +  \mathcal{L}_{unlabel},\\
	\mathcal{L}_{label} &= \frac{1}{|\mathcal{D}_0'|} \sum_{(x, p)\in \mathcal{D}_0'} \mathcal{L}_{cross}(p, \mathcal{M}'(x)),\\
	\mathcal{L}_{unlabel} &= \frac{1}{|\mathcal{D}_1'|} \sum_{(x, p)\in \mathcal{D}_1'} \|p - \mathcal{M}'(x)\|^2_2,
\end{aligned}
\end{equation}
where $\mathcal{L}_{cross}$ is the cross-entropy loss function, utilizing both label and unlabeled data.

\subsection{Local Adversarial Attack}\label{sub_local_attack}
With the substitute model obtained from the first step, the attacker can directly utilize the white-box adversarial attack to generate adversarial samples.

\begin{algorithm}[!t]
  \caption{Local Adversarial Attack (IPGD)}
  \label{protocol_white_box_attack}
  \begin{algorithmic}[1]
  
    \Require The target sample $x$ and its label $y$;
    optimization rate $\alpha$;
    tolerance perturbation $\epsilon$;
    maximum decay times $d_{max}$ 
    maximum iteration rounds $I$.
    
    \State Set $x_0^{adv} = x + noise$ where $noise$ is small random values and initialize $d = 0$.
    
    \For{$i \gets 0$ to $I$}
        \State $x_{i+1}^{adv}\gets \varphi_{\epsilon}({x_{i}^{adv} + \alpha sign(\nabla_{x} \mathcal{L}(\mathcal{M}'(x_{i}^{adv}),y))})$.
    
        \If{$\|x_i^{adv} - x_0^{adv}\|_\infty > \epsilon$ and $d < d_{max}$}
            \State Update $\alpha \gets decay(\alpha, i)$ and $d = d + 1$. 
            \State Back to the last iteration for optimization.
        \EndIf
        
    \EndFor
    
    \State \textbf{Return} the adversarial sample $x_{I}^{adv}$.
    
  \end{algorithmic}
\end{algorithm}

      

    







Intuitively, any white-box attack method can be leveraged.
In \sysname, we employ the Projected Gradient Descend (PGD)~\cite{madry2017towards}, the state-of-the-art white-box adversarial attack method.
Except for its superior performance, PGD avoids local minima through the random perturbation initialization method so that it can achieve a high transferability rate on the generated adversarial samples.
To enhance the attack success rate, \sysname further improves the current PGD convergence criterion as follows.

Given an arbitrary target input $x$,  we first initialize an initial adversarial sample $x_0^{adv}$ with a small noise perturbation.
\begin{equation}
x_0^{adv} = x + noise,
\end{equation}

Subsequently, we iteratively optimize the noises based on Eq.~\ref{eq_pgd} to find an optimal perturbation on $x$ that satisfies Definition~1.
\begin{equation}\label{eq_pgd}
x_{i+1}^{adv}=\varphi_{\epsilon}({x_{i}^{adv} + \alpha sign(\nabla_{x} \mathcal{L}(\mathcal{M}'(x_{i}^{adv}),y))}),
\end{equation}
where $x_{i+1}^{adv}$ is the adversarial sample generated at the $i_{th}$ iteration, $\alpha$ controls the perturbation optimization rate, $\mathcal{L}$ is the cross-entropy loss, $\varphi$ is the projection function $\|\cdot\|_\infty$ and $\epsilon$ is the maximum tolerance perturbation range. 

Finally, to improve the transferability of the generated adversarial examples, an intuitive way is to make the perturbation as ``obvious'' as possible.
In other words, with more perturbation, the generated adversarial samples will be more robust and generalizable, such that the attack model learned on the local substitute model can be better transferred to the target model.
An extreme example is that if we turn each pixel of a grey image to be $1.0$, the target model is hardly able to identify it like the original image.
Moreover, \sysname introduces the step rate (i.e., learning rate) decay mechanism into the attack procedure, as illustrated in Algorithm~\ref{protocol_white_box_attack}. 
By introducing the $decay(\cdot)$ function~\cite{robbins1951stochastic}, \sysname is able to optimize more on the possible perturbation directions that are less explored.



\section{Experiment}
\label{sec_experiments}

\begin{table*}[!ht]
\centering
\caption{Comparison of the untargeted adversarial attack performance on MNIST, Fashion-MNIST, and CIFAR-10.
}
\label{table_asr_query}
\begin{tabular}{@{}c|cc|cc@{}}
\toprule
\textbf{Datasets} & \multicolumn{2}{c|}{\textbf{MNIST}}       & \multicolumn{2}{c}{\textbf{CIFAR-10}}     \\ \midrule
\textbf{Method}   & \textbf{ASR} & \textbf{Query Number} & \textbf{ASR} & \textbf{Query Number} \\ \midrule
{SemiAdv}        &   91.29   &   1600   &  84.26    &  1600 \\
{ALA-IGS}         & 75.73  & 1600    & 61.43  & 6400    \\
{ALA-DF}         & 87.58  & 1600     & 72.53  & 6400    \\
{DaST-P}      & 53.99  & $>10^7$     & 29.10  & $>10^7$        \\
{DaST-L}      & 23.22  & $>10^7$     & 17.80  & $>10^7$        \\
{Pre-Trained} & 37.93  & $>10^7$     & 35.40  & $>10^7$        \\
{SignHunter}  & 91.25  & 10622      & 85.00  & 12510  \\
{BO-ADMM}    & 87.00  & 5210        & 84.10  & 4630   \\
{Bayes Attack}  & 90.35  & 2756      & 70.38  & 7588  \\
{Sign-OPT}  & 47.02  & 68236      & 31.87  & 67939  \\ \bottomrule
\end{tabular}%
\end{table*}

\begin{table*}[!ht]
\centering
\caption{Comparison of the targeted adversarial attack performance on MNIST, Fashion-MNIST, and CIFAR-10.}
\label{target_attack}
\begin{tabular}{@{}c|cc|cc|cc@{}}
\toprule
\textbf{Datasets} & \multicolumn{2}{c|}{\textbf{MNIST}} & \multicolumn{2}{c|}{\textbf{Fashion-MNIST}} & \multicolumn{2}{c}{\textbf{CIFAR-10}} \\ \midrule
\textbf{Method} & \textbf{ASR} & \textbf{QN} & \textbf{ASR} & \textbf{QN} & \textbf{ASR} & \textbf{QN} \\ \midrule
SemiAdv & 62.46 & 1600 & 58.89 & 1600 & 54.73 & 1600 \\
ALA-IGS & 51.58 & 1600 & 46.13 & 1600 & 34.39 & 1600 \\
ALA-DF & 52.41 & 1600 & 48.34 & 1600 & 37.47 & 1600 \\
DaST-P & 47.57 & $>10^7$ & 43.03 & $>10^7$ & 14.09 & $>10^7$ \\
Dast-L & 19.25 & $>10^7$ & 16.24 & $>10^7$ & 8.32 & $>10^7$ \\
Pre-Trained & 28.95 & $>10^7$ & 25.89 & $>10^7$ & 10.46 & $>10^7$ \\
Bayes Attack & 26.23 & 13003 & 24.87 & 13423 & 48.93 & 14915 \\
Sign-OPT & 2.41 & 97567 & 1.85 & 99423 & 3.5 & 93765 \\ \bottomrule
\end{tabular}
\end{table*}

\begin{table*}[!ht]
\centering
\caption{ASRs of different attack methods under different query numbers. QN specifies the query number.}
\label{table_white_box}
\begin{tabular}{@{}c|c|ccccc@{}}
\toprule
\textbf{Dataset} & \textbf{Attack} & \textbf{QN=100} & \textbf{QN=200} & \textbf{QN=400} & \textbf{QN=800} & \textbf{QN=1600} \\ \midrule
\multirow{4}{*}{MNIST} & FGSM & 67.80 & 71.23 & 73.23 & 75.37 & 79.70 \\
 & BIM & 73.61 & 75.61 & 76.95 & 80.28 & 83.51 \\
 & PGD & 77.32 & 81.41 & 83.21 & \textbf{86.62} & 89.01 \\
 & IPGD & \textbf{79.38} & \textbf{82.34} & \textbf{84.55} & 85.92 & \textbf{91.29} \\ \midrule
\multirow{4}{*}{Fashion-MNIST} & FGSM & 63.02 & 67.08 & 68.75 & 69.26 & 75.65 \\
 & BIM & 69.14 & 69.23 & 69.58 & 76.30 & 81.52 \\
 & PGD & 71.52 & 73.77 & 76.09 & 79.74 & 82.57 \\
 & IPGD & \textbf{74.17} & \textbf{76.06} & \textbf{78.49} & \textbf{82.12} & \textbf{84.91} \\ \midrule
\multirow{4}{*}{CIFAR-10} & FGSM & 30.57 & 40.38 & 47.87 & 51.67 & 55.60 \\
 & BIM & 42.15 & 56.63 & 68.68 & 72.42 & 75.60 \\
 & PGD & 45.88 & 59.54 & 72.08 & 75.72 & 81.80 \\
 & IPGD & \textbf{47.47} & \textbf{62.05} & \textbf{73.75} & \textbf{78.87} & \textbf{84.26} \\ \bottomrule
\end{tabular}
\end{table*}


\begin{figure*}[!ht]
\centering
\subfigure[MNIST]{
    \begin{minipage}[t]{0.32\linewidth}\centering
        \includegraphics[scale=0.2]{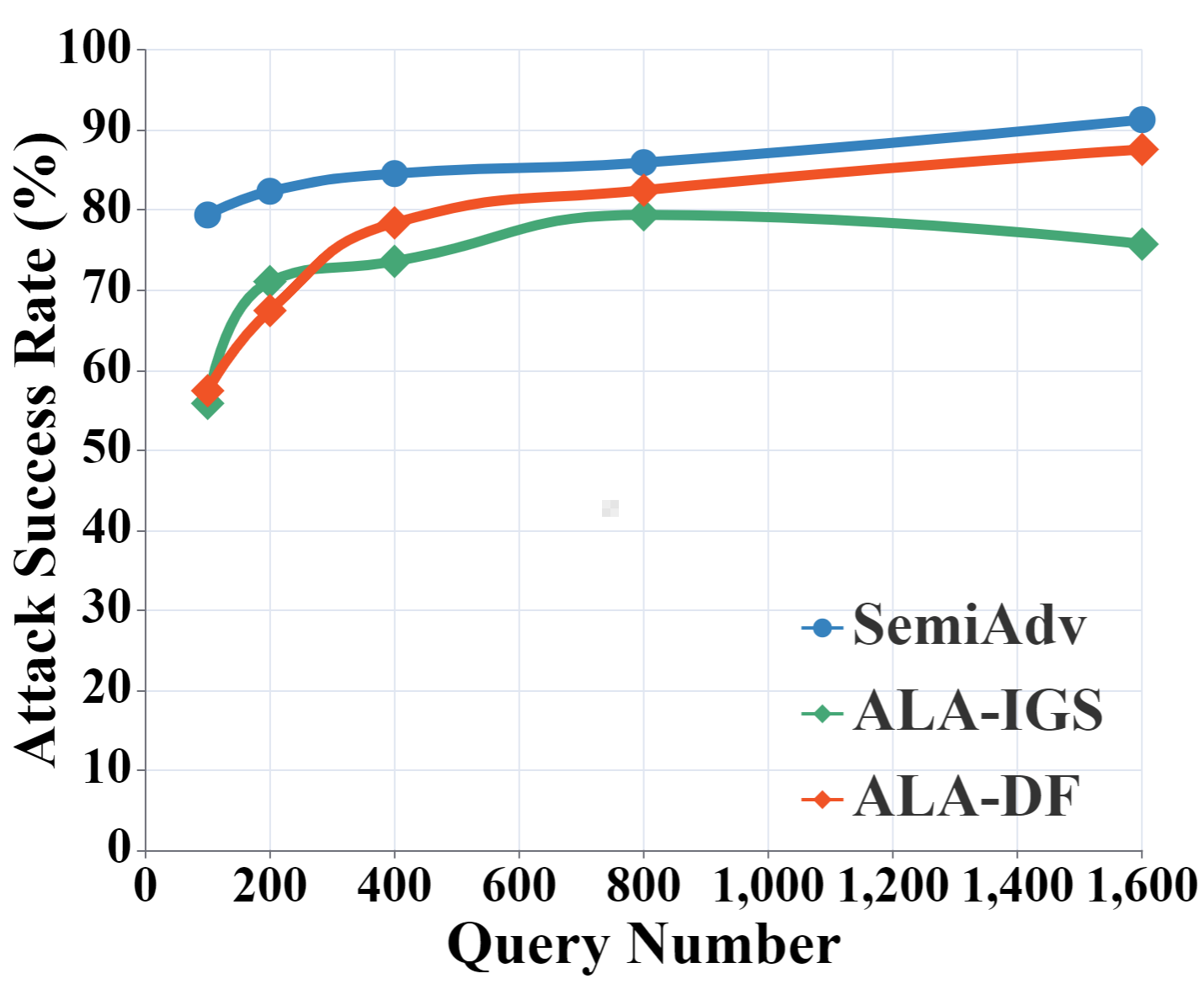}
    \end{minipage}
}
\subfigure[Fashion-MNIST]{
    \begin{minipage}[t]{0.32\linewidth}\centering
        \includegraphics[scale=0.2]{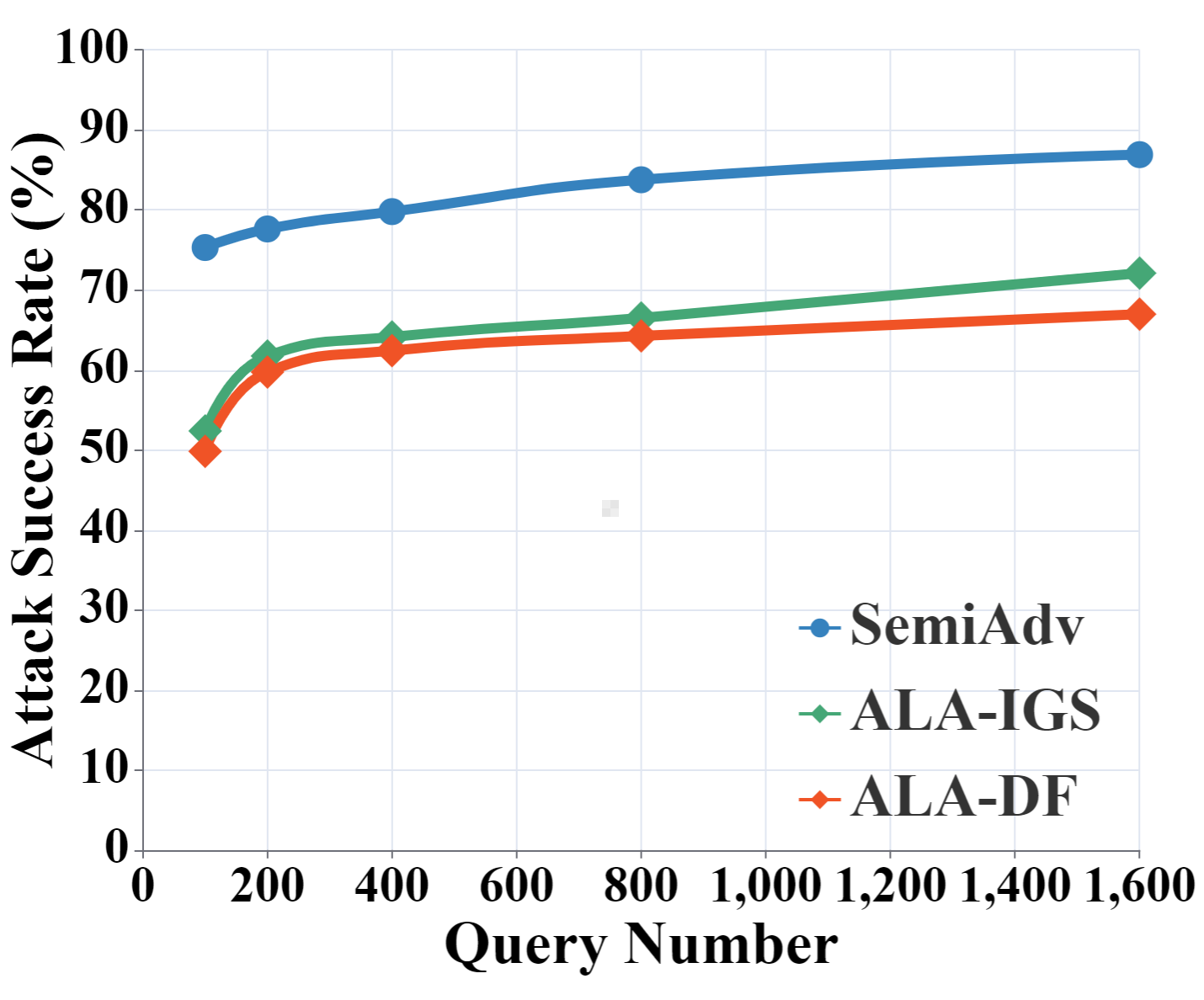}
    \end{minipage}
}
\subfigure[CIFAR-10]{
    \begin{minipage}[t]{0.32\linewidth}\centering
        \includegraphics[scale=0.2]{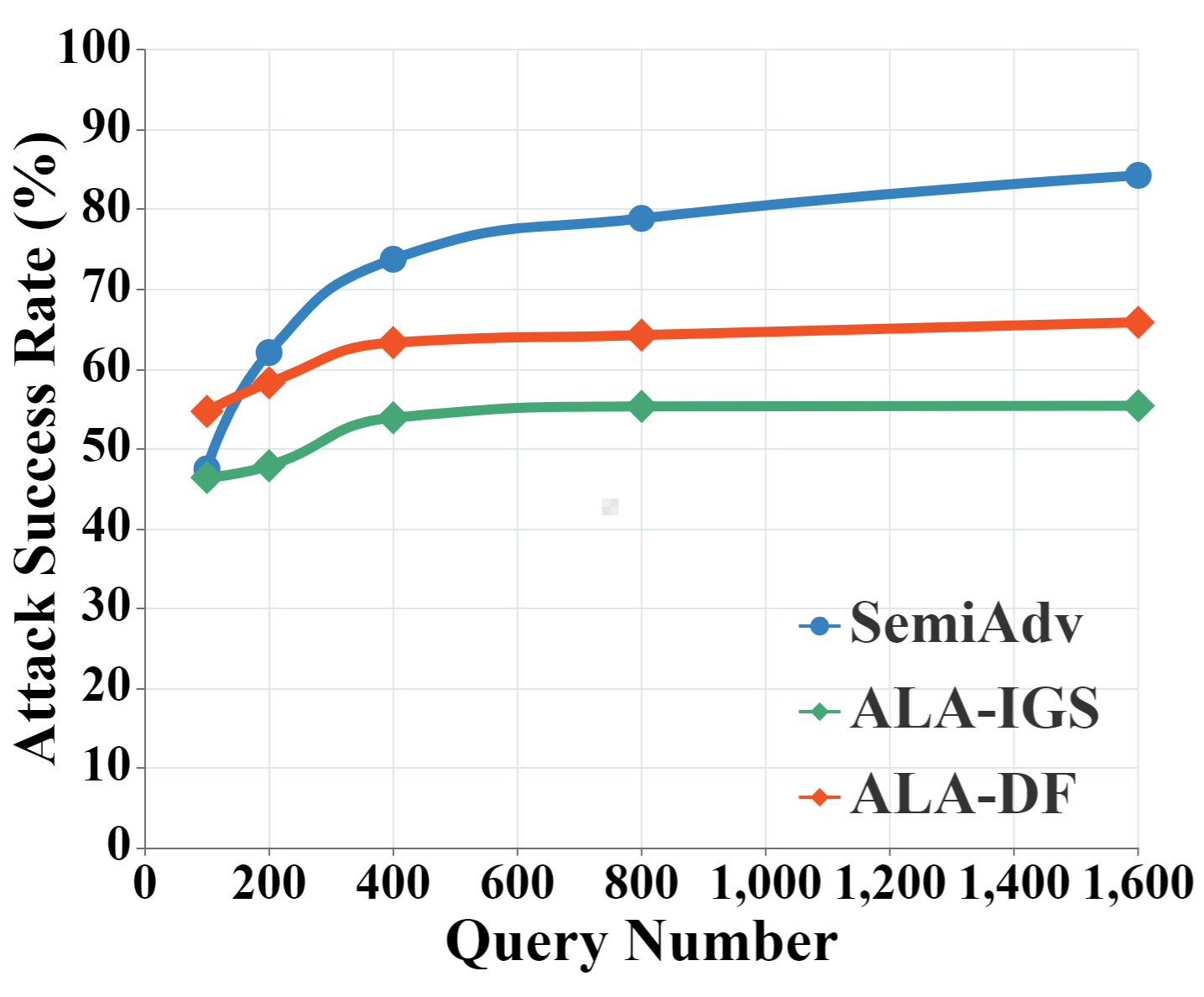}
    \end{minipage}
}
\caption{Performance of different methods under different query numbers on MNIST, Fashion-MNIST, and CIFIAR-10.}
\label{method_comparison_fig}
\end{figure*}

\begin{table*}[!ht]
\centering
\caption{Comparison of similarity (measured by accuracy \%) between the substitute model and target model.}
\label{compare_similarity}
\begin{tabular}{@{}c|c|ccccc@{}}
\toprule
\multicolumn{2}{c|}{\textbf{Query Number}} & \textbf{100} & \textbf{200} & \textbf{400} & \textbf{800} & \textbf{1600} \\ \midrule
\multirow{3}{*}{{MNIST}} & {SemiAdv} & \textbf{95.21} & \textbf{96.71} & \textbf{97.16} & \textbf{97.58} & \textbf{98.61} \\
 & {ALA-IGS} & 47.64 & 56.84 & 76.48 & 76.84 & 79.62 \\
 & {ALA-DF} & 40.09 & 48.94 & 74.15 & 76.79 & 86.28 \\ \midrule
\multirow{3}{*}{{\begin{tabular}[c]{@{}c@{}}Fashion-\\ MNIST\end{tabular}}} & {SemiAdv} & \textbf{93.94} & \textbf{95.69} & \textbf{96.18} & \textbf{96.54} & \textbf{97.72} \\
 & {ALA-IGS} & 45.91 & 56.20 & 76.13 & 76.08 & 76.63 \\
 & {ALA-DF} & 38.59 & 46.89 & 72.45 & 76.47 & 83.53 \\ \midrule
\multirow{3}{*}{{CIFAR-10}} & {SemiAdv} & 51.68 & \textbf{65.00} & \textbf{74.26} & \textbf{79.80} & \textbf{84.87} \\
 & {ALA-IGS} & 53.51 & 52.71 & 57.13 & 55.82 & 63.64 \\
 & {ALA-DF} & \textbf{53.61} & 57.13 & 64.71 & 74.73 & 77.47 \\ \bottomrule
\end{tabular}
\end{table*}

\begin{table*}[!ht]
\centering
\caption{Attack effectiveness on substitute models against different target models with varying query numbers on MNIST.}
\label{table_asr_arch_mnist}
\begin{tabular}{@{}c|c|cccc|cccc@{}}
\toprule
\multicolumn{2}{c|}{\textbf{Substitute Model}} & \multicolumn{4}{c|}{\textbf{EfficientNet}} & \multicolumn{4}{c}{\textbf{Wide ResNet28}} \\ \midrule
\multicolumn{2}{c|}{\textbf{Target Model}} & \textbf{GoogleNet} & \textbf{MobileNet} & \textbf{PreActResNet18} & \textbf{Average} & \textbf{GoogleNet} & \textbf{MobileNet} & \textbf{PreActResNet18} & \textbf{Average} \\ \midrule
\multirow{5}{*}{Query Number} & 100 & 83.01 & 83.78 & 82.95 & 83.25 & 83.06 & 71.33 & 72.15 & 75.51 \\
 & 200 & 83.87 & 88.13 & 87.00 & 86.33 & 85.32 & 74.98 & 74.75 & 78.35 \\
 & 400 & 87.05 & 88.89 & 88.06 & 88.00 & 86.72 & 79.20 & 77.42 & 81.11 \\
 & 800 & 90.99 & 85.35 & 85.25 & 87.20 & 87.08 & 83.90 & 82.93 & 84.64 \\
 & 1600 & 91.02 & 92.65 & 91.34 & 91.67 & 89.37 & 95.86 & 87.50 & 90.91 \\ \bottomrule
\end{tabular}
\end{table*}

\begin{figure*}[h]
\centering
\subfigure[MNIST]{
    \begin{minipage}[t]{0.26\linewidth}\centering
        \includegraphics[scale=0.17]{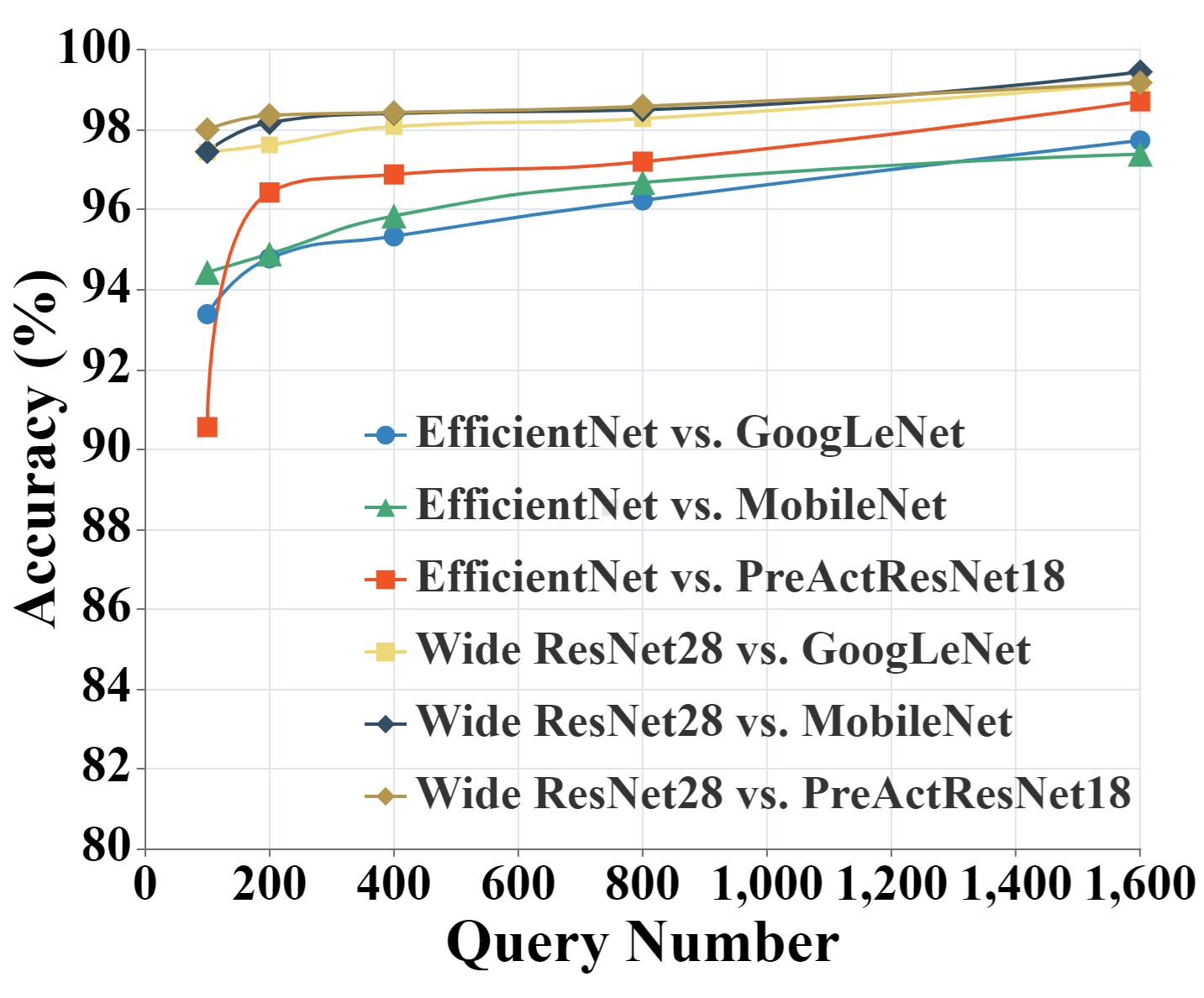}
    \end{minipage}
}
\hfill
\subfigure[Fashion-MNIST]{
    \begin{minipage}[t]{0.26\linewidth}\centering
        \includegraphics[scale=0.17]{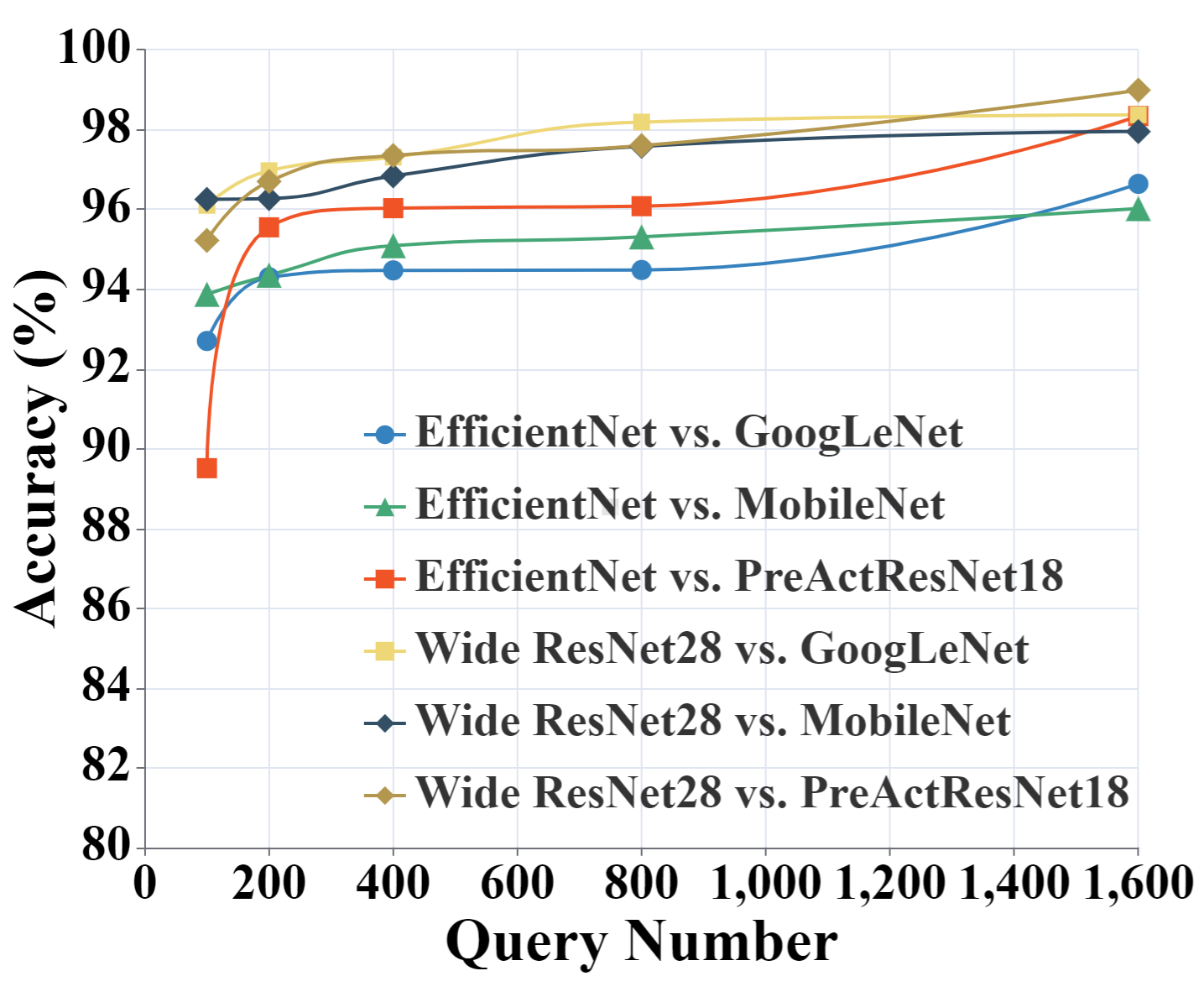}
    \end{minipage}
}
\hfill
\subfigure[CIFAR-10]{
    \begin{minipage}[t]{0.26\linewidth}\centering
        \includegraphics[scale=0.17]{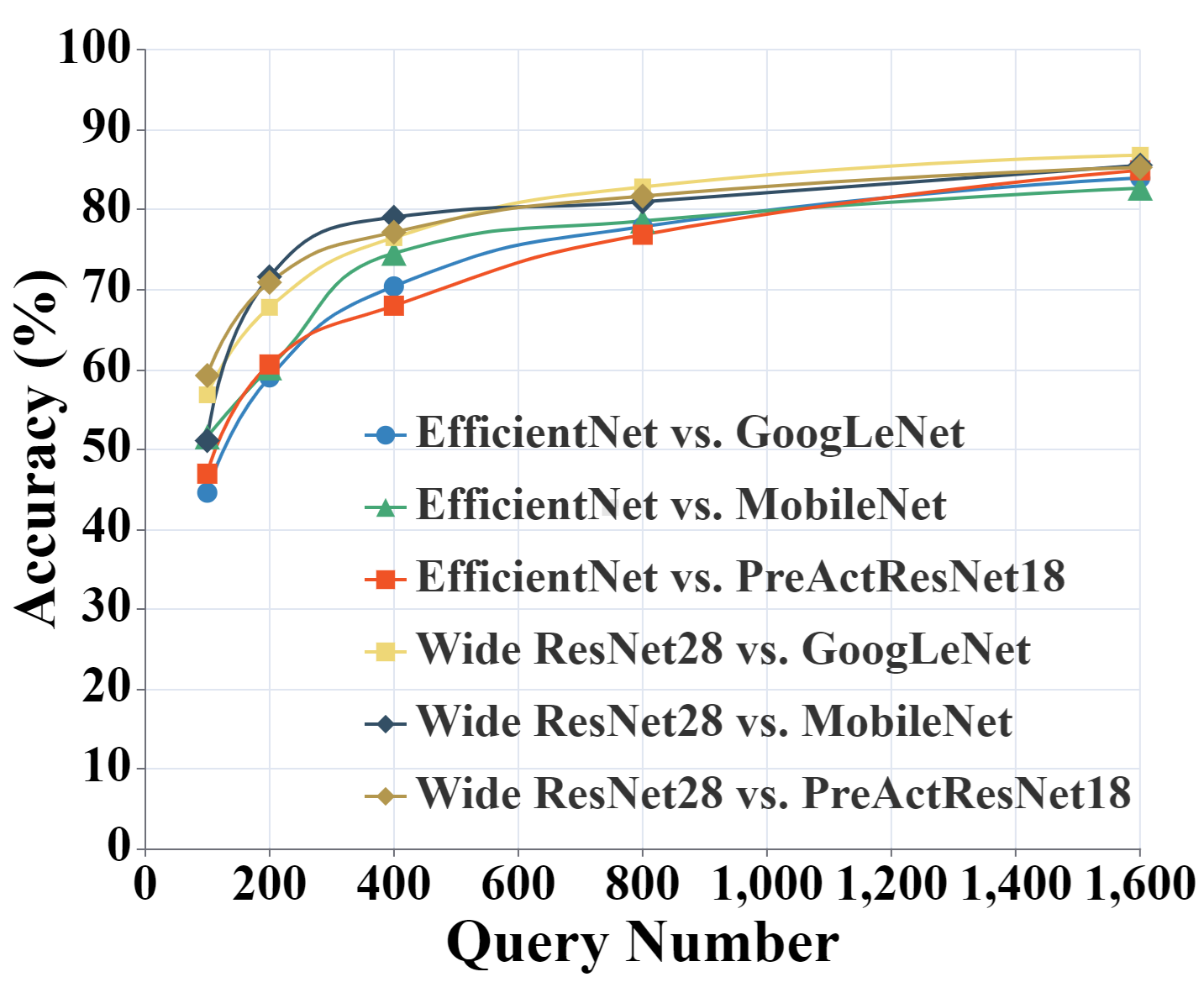}
    \end{minipage}
}
\caption{Accuracy of different substitute model architectures under query numbers ($\in \{100,200,400,800,1600\}$) in
 three benchmark datasets MNIST, Fashion-MNIST, and 
CIFAR-10.}
\label{fig_arch_cifar}
\end{figure*}

In this section, we conduct extensive experiments across several benchmark datasets and various model architectures to evaluate the performance of \sysname against the state-of-the-art black-box adversarial attack baselines.
The experimental results show that \sysname significantly outperforms these baselines, especially on the number of required query access.

\subsection{Experimental Setup}



\textbf{Dataset.} 
We validate \sysname on three benchmark datasets, namely MNIST, Fashion-MNIST, and CIFAR10.
Specifically, the training sets of the two datasets are randomly partitioned into halves of the same size. 
The first half is used to train the target model, while the other part is treated as unlabeled data to fit the substitute model.

\textbf{Network Architecture.} 
To evaluate the robustness of our attacks, we experiment with five widely used network architectures on \sysname. 
For the target model, three network structures are tested, i.e., MobileNet, GoogLeNet, and PreActResNet18. 
For the substitute model, we choose Wide ResNet-28 and EfficientNet. 

\textbf{Black-box model training details.}
Black-box models (MobileNet, GoogLeNet, and PreActResNet18) were trained with 20 epochs for MNIST and Fashion-MNIST and 100 epochs for CIFAR-10 and CIFAR-100, over the corresponding training set.
Random crop and random horizontal flip were used as data augmentation for CIFAR-10 and CIFAR-100, while vanilla data processing is used in MNIST and Fashion-MNIST.
Throughout the training process of black-box models, we adopted momentum optimizer with 0.1 learning rate, 0.9 momentum factor, and $5 \times 10^{-4}$ weight decay ($L_2$ penalty) and cosine annealing learning rate scheduler with 200 maximum number of iterations.
Besides, cross entropy loss function is adopted, and the black-box models with the best accuracy are saved for attacking during the training process.

\textbf{Substitute model training details.}
We set learning rate (Adam optimizer) to 0.004, training iteration to 256, $\alpha$ to 0.75, $\lambda_{\upsilon}$ to 75, T to 0.5, decay rate to 0.999, and iterations of each epoch to 256.
The epoch was set to 20 in MNIST and Fashion-MNIST and 100 in CIFAR-10 and CIFAR-100.

\textbf{Baselines.} 
Seven state-of-the-art black-box adversarial attacks are considered here, including
Active Learning Attack (ALA)~\cite{activelearning}, Data-Free Attack (DaST)~\cite{DaST}, Pre-Trained Attack (Pre-Trained)~\cite{DaST}, SignHunter~\cite{signbits},  BO-ADMM~\cite{gradientfree},Bayes Attack~\cite{bayes_attack}, and Sign-OPT~\cite{sign_opt}.
The attacks cover the two mainstream black-box attack directions mentioned before.
Among them, we introduce two variants of ALA, i.e., Iterative Gradient Sign (ALA-IGS)~\cite{BIM} and DeepFool (ALA-DF)~\cite{DeepFool}.
For DaST and BO-ADMM, we implement them in both the label-only and probability-only versions, abbreviated as DaST-L, DaST-P, and DBO-ADMM, SBO-ADMM, respectively.
Sign-OPT and Bayes Attack are currently state-of-the-art under label-only black-box scenario.
Here, label-only means the attacker can only access the label outputs of the target model and probability-only indicates the attacker can get the prediction scores of the target model.
We also implement a naive DaST based attack method, i.e., Pre-Trained, which directly uses a pre-trained network as the substitute model without fine-tuning. 


\textbf{Others.}
By default, every experimental result is the average evaluation result of the above five network architectures.
The query budgets for attacks are set to \{100, 200, 400, 800, 1600\}.
Moreover, we evaluate the attack performance of all attacks
with same perturbation constraint of \cite{DaST}.
Three types of metrics are mainly considered, including the attack success rate (ASR), query number, and similarity.
\textit{ASR} reflects the misclassification rate of the target model on the generated adversarial samples.
\textit{Query number} indicates the upper bound on the number of queries these methods can perform for adversarial sample generation.
\textit{Similarity} refers to the performance similarity between the target model and the substitute model.

\begin{table*}[!ht]
\centering
\caption{Attack effectiveness on substitute models against different target models with varying query numbers on Fashion-MNIST.}
\label{table_asr_arch_fashion}
\begin{tabular}{@{}c|c|cccc|cccc@{}}
\toprule
\multicolumn{2}{c|}{\textbf{Substitute Model}} & \multicolumn{4}{c|}{\textbf{EfficientNet}} & \multicolumn{4}{c}{\textbf{Wide ResNet28}} \\ \midrule
\multicolumn{2}{c|}{\textbf{Target Model}} & \textbf{GoogleNet} & \textbf{MobileNet} & \textbf{PreActResNet18} & \textbf{Average} & \textbf{GoogleNet} & \textbf{MobileNet} & \textbf{PreActResNet18} & \textbf{Average} \\ \midrule
\multirow{5}{*}{Query Number} & 100 & 75.87 & 76.90 & 77.30 & 76.69 & 78.13 & 65.96 & 70.83 & 71.64 \\
 & 200 & 77.36 & 77.56 & 83.80 & 79.57 & 80.09 & 66.40 & 71.14 & 72.54 \\
 & 400 & 78.67 & 80.48 & 83.82 & 80.99 & 81.36 & 74.09 & 72.54 & 76.00 \\
 & 800 & 84.41 & 84.96 & 84.34 & 84.57 & 84.12 & 75.39 & 79.47 & 79.66 \\
 & 1600 & 85.77 & 85.50 & 85.93 & 85.73 & 85.18 & 83.69 & 83.38 & 84.08 \\ \bottomrule
\end{tabular}
\end{table*}

\begin{table*}[!ht]
\centering
\caption{Attack effectiveness on substitute models against different target models with varying query numbers on CIFAR-10.}
\label{table_asr_arch_cifar}
\begin{tabular}{@{}c|c|cccc|cccc@{}}
\toprule
\multicolumn{2}{c|}{\textbf{Substitute Model}} & \multicolumn{4}{c|}{\textbf{EfficientNet}} & \multicolumn{4}{c}{\textbf{Wide ResNet28}} \\ \midrule
\multicolumn{2}{c|}{\textbf{Target Model}} & \textbf{GoogleNet} & \textbf{MobileNet} & \textbf{PreActResNet18} & \textbf{Average} & \textbf{GoogleNet} & \textbf{MobileNet} & \textbf{PreActResNet18} & \textbf{Average} \\ \midrule
\multirow{5}{*}{Query Number} & 100 & 48.90 & 61.29 & 58.60 & 56.26 & 35.10 & 42.60 & 38.30 & 38.67 \\
 & 200 & 53.40 & 65.03 & 64.10 & 60.84 & 63.30 & 64.86 & 61.60 & 63.25 \\
 & 400 & 69.70 & 75.78 & 74.30 & 73.26 & 72.70 & 75.63 & 74.40 & 74.24 \\
 & 800 & 73.70 & 81.25 & 78.50 & 77.82 & 77.20 & 83.36 & 79.20 & 79.92 \\
 & 1600 & 82.40 & 83.67 & 82.90 & 82.99 & 83.70 & 87.46 & 85.40 & 85.52 \\ \bottomrule
\end{tabular}
\end{table*}

\subsection{Comparison with State-of-The-Art}
Table~\ref{table_asr_query}, Table~\ref{target_attack}, and Table~\ref{table_white_box} summarize the main comparison results.
In the experiments, the metrics are all evaluated on the whole testing sets.
Overall, \sysname outperforms all the previous methods on both ASR and query number.
Next, we present the detailed comparison results from five different aspects.

\paragraph{\sysname vs. ALA \& DaST} 
For the \textit{transferability-based black-box adversarial attacks}, ALA and its variants achieve the most competitive performance.
However, as shown in Table~\ref{table_asr_query}, \sysname still yields a gain of 12\% on ASR with same queries budget in CIFAR-10, compared with ALA.
Fig.~\ref{method_comparison_fig} also shows that \sysname consistently outperforms ALA even when the query number is very small.
DaST and Pre-Trained rely on generative adversarial networks to synthesize data for substitute model training.
As the synthetic data generation is hard to craft and needs massive queries, DaST and Pre-Trained perform poorer than \sysname, especially for the query number.

\paragraph{\sysname vs. SignHunter \& BO-ADMM} 
Compared to the gradient estimation methods, \sysname shows higher overall performance.
Although the ASR of SignHunter is similar to \sysname, SignHunter needs a large number of queries to produce effective adversarial examples.
Specifically, \sysname only needs 1600 queries on MNIST and CIFAR-10 to launch attacks while SignHunter requires about ten times more queries.
Moreover, with fewer queries, \sysname achieves 4.29\%, 0.88\%, and 0.16\% performance improvement w.r.t. ASR over BO-ADMM on MNIST, Fashion-MNIST and CIFAR-10, respectively.

\paragraph{\sysname vs. Sign-OPT \& Bayes Attack} 
Compared to two advanced query-based black-box adversarial attacks Sign-OPT and Bayes Attack, \sysname still dominates on both ASR and queries, particularly on sophisticated dataset (CIFAR-10).
It is illustrated in Table \ref{table_asr_query} that \sysname can obtain high ASR while only require one-fourth to one-second query budgets of Bayes-Attack, suggesting the superior performance of \sysname.

\paragraph{\sysname vs. Target Attack} 
Target attack is a more challenging and practical attack than untarget attack in this area, and we hence examine the performance of \sysname and baselines on target attack \ref{target_attack} to further validate the effectiveness of \sysname.
Similar to the phenomenon in Table \ref{table_asr_query}, \sysname consistently obtains higher ASR with fewer query numbers, which suggests \sysname is a better black-box attack method compared to others, no matter target or untarget settings.

\subsection{The Evaluation of \sysname in Different Conditions}


\paragraph{\sysname vs. White-Box} 
First, one of the key steps of \sysname is to attack the locally estimated substitute model with the white-box adversarial attacks.
Here, we denote our attack method proposed in Algorithm~\ref{protocol_white_box_attack} as IPGD.
The first observation is that PGD achieves the best performance among all previous white-box attack methods.
Furthermore, IPGD outperforms the pure PGD attack given the same query number, which proves the effectiveness of our method.
A second observation is that FGSM and BIM perform quite poorly over adversarial transferability, which indicates that they are inappropriate to be applied in transferability based black-box adversarial attacks.
As shown in Table~\ref{table_white_box}, the FGSM attack with 1600 queries achieves 55.60\% ASR, while, IPGD reaches 62.05\% ASR with just 200 queries on CIFAR-10.

\paragraph{\sysname vs. Similarity}
As above-mentioned, the performance of transferability based attacks has a positive correlation to the similarity between the substitute model and the target model.
Therefore, except for ASR, we also compare the similarity metric between \sysname and ALA, as shown in Table~\ref{compare_similarity}.
Here, we omit DaST related results because, from Table~\ref{table_asr_query}, its performance is not competitive at all. 
Moreover, the gradient estimation based methods are also omitted because these methods do not involve substitute models in the attack procedure.
From the results, \sysname convergences faster than ALA, which is reflected by closer accuracy to the target model on the testing data, which justifies the fact why \sysname achieves better performance on ASR.
Moreover, combined with ASR related experiment results, we discover that as the substitute model performance increases, the ASR of \sysname increases in a similar trend.

\paragraph{\sysname vs. Query Number}
Next, we experiment with the performance of \sysname under different query numbers.
All figures and tables of these experiments contain the trade-off evaluation between ASR/similarity and the number of queries.
As we can see, the performance of \sysname grows positively with the query number.
Besides, the first few hundred of labels contribute a lot to the effectiveness of \sysname attack. 
The phenomenon hints that \sysname is more practical for real-world applications, as fewer adversarial queries are more likely for the attacker to hide the attack pattern.

\paragraph{\sysname vs. Network Architecture}
According to the black-box setting, the attacker should have no prior knowledge of the network architecture of the target model.
Thus, we further investigate the effect of different network architectures on \sysname performance.
These results are presented in Table~\ref{table_asr_arch_mnist}, Table~\ref{table_asr_arch_fashion}, Table~\ref{table_asr_arch_cifar} and Figure~\ref{fig_arch_cifar}.
In the experiments, \sysname demonstrates to be more robust to the change in architectures.
Specifically, no matter which types of model architecture are chosen, \sysname can always attain almost 80\% ASR and maintain high similarity of the substitute model to the target model, which further verifies the practicality of \sysname in practice.

\section{Conclusion}
\label{sec_conclusion}
In this research, we focused on investigating the vulnerability of deep neural networks against adversarial attacks.
Compared with attack settings adopted by prior works, a more practical black-box attack setting was considered in this paper where attackers are allowed to access the output of the target model w.r.t. each instance once.
Furthermore, we presented \sysname to efficiently launch adversarial attacks in this challenging setting, which is a novel and query-efficient black-box transfer-based adversarial attack method with unlabeled data.
\sysname dramatically reduced the query budget for substitute model training by introducing an optimized semi-supervised learning algorithm.
Besides, we highlight that, this was the first method that could launch black-box adversarial attacks with only unlabeled raw data. 
Such unlabeled data can easily be found in many public resources for attackers.
Moreover, we modified the local white-box attack method to improve the transferability of produced adversarial samples, which can significantly enhance the attack effectiveness of \sysname further.
Finally, we conducted extensive experiments across various model architectures in multiple benchmark datasets to exhaustively examine the attack effectiveness of \sysname.
Wherein, the experimental results showed that \sysname outperformed state-of-the-art methods in terms of both attack success rates and the required query access, suggesting the superior attack performance of \sysname compared with state-of-the-art baselines. 

\section{Ethics Statement}

This paper designs an adversarial attack method. While the method seems harmful, it is believed that the benefits of publishing \sysname outweigh the potential harms. It is better to expose the blind spots of models as soon as feasible because doing so can alert deployers to be aware of potential threats and greatly encourage AI community to design corresponding defense strategies.

\clearpage





\bibliography{references}

\end{document}